\pdfoutput=1
\documentclass[twocolumn]{aastex631}

\pdfpageattr{/Group << /S /Transparency /I true /CS /DeviceRGB>>}

\usepackage{enumitem}
\usepackage{amsmath}
\usepackage{graphicx}
\usepackage{tikz}
\shorttitle{AASTeX v6.3.1 Sample article}
\shortauthors{Cox et al.}
\graphicspath{{./}{figures/}}

\begin{document}

\title{X-ray Absorption Variability in NGC\,1142: Another Constraint on the Nature of the Torus/Broad-Line Region in Active Galactic Nuclei}

\author[0000-0003-2287-0325]{Isaiah S. Cox}
\affiliation{Department of Physics and Astronomy, Clemson University, Clemson, SC, 29634, USA}

\author[0000-0003-3638-8943]{Núria Torres-Alb\`a}
\affiliation{Department of Astronomy, University of Virginia, 
P.O. Box 400325, Charlottesville, VA 22904, USA}

\author[0000-0001-5544-0749]{Stefano Marchesi}
\affiliation{Dipartimento di Fisica e Astronomia, Università degli Studi di Bologna, via Gobetti 93/2, 40129 Bologna, Italy}
\affiliation{Department of Physics and Astronomy, Clemson University, 
Clemson, SC, 29634, USA}
\affiliation{INAF-Osservatorio Astronomico di Bologna, Via Piero Gobetti, 93/3, I-40129, Bologna, Italy}

\author[0000-0002-9719-8740]{Vittoria E. Gianolli}
\affiliation{Department of Physics and Astronomy, Clemson University, 
Clemson, SC, 29634, USA}
\affiliation{INAF-Osservatorio Astronomico di Brera, Via Brera 28, 20121 Milano, Italy}

\author[0000-0002-7791-3671]{Xiurui Zhao}
\affiliation{Department of Astronomy, University of Illinois at Urbana-Champaign, Urbana, IL 61801, USA}

\author[0000-0002-6584-1703]{Marco Ajello}
\affiliation{Department of Physics and Astronomy, Clemson University, 
Clemson, SC, 29634, USA}

\author[0000-0002-7825-1526]{Indrani Pal}
\affiliation{Department of Physics and Astronomy, Clemson University, 
Clemson, SC, 29634, USA}

\author[0000-0001-6412-2312]{Andrealuna Pizzetti}
\affiliation{European Southern Observatory, Alonso de Córdova 3107, Casilla 19, Santiago 19001, Chile}






\begin{abstract}

The physics, geometry, and kinematics of the gas and dust surrounding the central engine of active galactic nuclei (AGN) are not well understood. In some sources, this material obscures the coronal X-ray emission, allowing its line-of-sight column density ($N_H$) to be measured. Many sources display $N_H$ variability across multiple epochs. Studying this variability can provide information about the distribution and dynamics of the obscuring material. Particularly, previous works have shown that the probability of observing $N_H$ variability increases with the number of observations available. Additionally, the observed magnitudes of the $N_H$ variability depend on the timescale between observations. In this work, we present the $N_H$ variability analysis of NGC\,1142, which has a total of nine X-ray observations over $\sim20$\,years, including five from a NuSTAR+XMM-Newton monitoring campaign performed in 2024-2025 and reported here for the first time. We use the physically-motivated torus models \texttt{UXCLUMPY}, \texttt{XSKIRTor}, and \texttt{RXTorusD} to model the spectra. Our results confirm the previously observed trends, and we compare these trends to simulated obscuration curves assuming different locations and distributions of obscuring clouds.

\end{abstract}



\section{Introduction} \label{sec:intro}

Active galactic nuclei (AGN) are supermassive black holes (SMBHs) at the centers of galaxies that are actively accreting matter and radiating at all wavelengths. In particular, the primary X-ray emission is produced by the comptonization of accretion disk photons in a region called the corona \citep[e.g.][]{Haardt93}. Many specific details of the corona remain unknown, however, techniques like reverberation mapping have established that it is small ($\lesssim15\,r_g$) and located very close to the SMBH \citep[e.g.][]{Fabian09,Kara13,Cackett14,Emmanoulopoulos14,Uttley14,Fabian15,Kara16,Chainakun19}. This emission is in the form of a powerlaw with a typical photon index $\sim$1.6--2.0 and a high-energy cutoff of a few hundred keV \citep[e.g.][]{Nandra94,Reeves2000,Shemmer08,Zappacosta18,Ricci18}. The powerlaw emission can be significantly reprocessed by the disk and gas/dust at radii extending out to several parsecs \citep[e.g.][]{Lightman1988,Pounds1990,Turner09,Brightman11,buchner14bxa}. 

This reprocessing is often associated with the structure commonly referred to as the `torus,' which has historically been invoked to explain the differences between Type 1 and Type 2 AGN as an effect of inclination angle \citep[e.g.][]{antonucci_unified_1993,urry95}. In Type 2 AGN, this torus intercepts the observer's line of sight often resulting in a significantly harder observed spectrum caused by photoelectric absorption \citep[e.g.][]{DellaCeca08,Burlon11,Vasudevan13,Ricci15,Lanzuisi18,peca_x-ray_2021,NTA21}. The amount of absorption depends on the amount of material integrated along the line of sight, which is quantified by the hydrogen column density, $N_H$. The $N_H$ can be very precisely determined by measuring the low-energy cutoff in the primary powerlaw. 

Since the corona is compact and located close to the SMBH, the $N_H$ as measured by the X-ray spectrum is a very powerful probe into the nature of the torus and broad-line region (BLR) as it provides a relatively localized measurement of material that is otherwise spatially unresolved, except for a few very nearby sources using infrared interferometry \citep[e.g.][]{Jaffe04,Tristram07,Raban09,Burtscher13}. Various observations and theoretical considerations point to a scenario where the obscuring material being inhomogeneous \citep[e.g.][]{nenkova02,nenkova08,gandhi09,stalevski12,alonso16,garcia16}. Outflows and winds from the accretion disk have been proposed as mechanisms to produce and sustain this clumpy material in a geometrically thick distribution \citep[e.g.][]{Konigl94,Elitzur06,Wada12,Schartmann14,Honig19}.

Since the $N_H$ probes a sliver of the dynamic and inhomogeneous material, variability in $N_H$ can provide an independent constraint on the distribution of the material in the BLR/torus. Many sources display variability in $N_H$ on timescales ranging from less than a day \citep[e.g.][]{Elvis2004,Risaliti2005,Puccetti2007,Bianchi09,Maiolino2010,Sanfrutos13,Miniutti2014}, to weeks or months \citep[e.g.][]{Lamer03,Marinucci13,Jana22,NTA25} and to more than a year \citep[e.g.][]{Markowitz2014,hernandez2015x, Laha2020,NTA23,pizzetti_multi-epoch_2022,Pizzetti2025}. This obscuration variability is most commonly attributed to eclipsing events due to clouds in the BLR and attempts have been made to use these eclipses to constrain the size, location, and even the shape of the clouds \citep[e.g.][]{Risaliti2002,Risaliti2005,maiolino_comets_2010,pizzetti_multi-epoch_2022,marchesi_compton-thick_2022,NTA25}. However, in reality, a complex dance of tangled outflowing/inflowing/orbiting material likely contribute to the observed absorption variability \citep[e.g.][]{Kaastra2014,mehdipour2017,Kara2021}.

Over the past few years, our group\footnote{https://science.clemson.edu/ctagn/} has been working on obtaining $N_H$ variability measurements for a large sample of sources that were originally determined to require variability in either flux or absorption based on a simultaneous fit to non-simultaneous NuSTAR and soft X-ray data from \cite{Zhao21}. The results for 28 sources have been published in \cite{pizzetti_multi-epoch_2022,NTA23,Pizzetti2025,Sengupta25,NTA25}. The results for another 11 sources that did not require variability in \cite{Zhao21} was published in \cite{Gianolli2026}. 

Interestingly, a significant fraction ($\sim50$\,\%) of the sources in the variable sample turn out to have no evidence of $N_H$ variability. This is also seen in other works \citep[e.g.][]{hernandez2015x,Laha2020}. Similarly, from the `non-variable' sample, a significant fraction of sources turn out to display $N_H$ variability. An important result from these works is that the probability of observing variability depends on the number of observations considered \citep[e.g.][]{Cox26}. More importantly, it appears that this probability increases at a specific rate. Measuring this rate can constrain physical properties of the obscuring material, as we discuss in this work. 

Specifically, in this work, we provide $N_H$ measurements for 9 X-ray observations of NGC\,1142, a Seyfert 2 galaxy located at a distance of $d\sim120$\,Mpc $(z=0.0287)$. The black hole mass ranges in the literature from $M_{\rm BH}=10^{8.2-9.4}$\,M$_{\odot}$ \citep{Winter09,deRosa12,Lubinski16,Kawamuro16,Kammoun20,Osorio_Clavijo22,Akylas22} and the Eddington ratio ranges from $\log \lambda_{\text{Edd}}\sim$-2.5 to -1.6 \citep{Lubinski16,Kawamuro16,Kammoun20,Panessa22}. Four observations of NGC\,1142 were previously analyzed in \cite{Pizzetti2025} and $N_H$ variability was found. $N_H$ variability was also seen in \cite{Kammoun20}. We extend their analysis by including 5 more NuSTAR observations (4 of which have simultaneous XMM-Newton data) taken between 2024 and 2025 over a timespan of five months.

In Section \ref{sec:data}, we describe the reduction of the XMM-Newton, NuSTAR, and Suzaku data. In Section \ref{sec:models}, we describe the models used and the fitting procedures employed. We show the spectra and best-fit parameters of our models in Section \ref{sec:results}. In Section \ref{sec:discussion}, we compare the results of the three models and identify trends in the probability of detecting variability as well as the magnitude of variability. In addition, we introduce a simple model to simulate absorption variability and use our results to constrain the model. We summarize our results in Section \ref{sec:conclusions}. Throughout the paper, we use a standard cosmology, with $H_0=70$\,km\,s$^{-1}$\,Mpc$^{-1}$, $\Omega_m=0.27$, and $\Omega_{\lambda}=0.73$. All reported errors are at the 90\,\% confidence level unless stated otherwise.


\section{Data} \label{sec:data}

We downloaded and reprocessed all available archival XMM-Newton, NuSTAR, and Suzaku data for NGC\,1142 in addition to the newly obtained XMM-Newton+NuSTAR monitoring campaign. In total, there are nine separate epochs with available X-ray data, four of which have simultaneous XMM-Newton+NuSTAR data. The observation details are shown in Table \ref{tab:obs}.

\begin{deluxetable}{ccccc}
\tabletypesize{\scriptsize}

\tablecaption{The observations used in this work. \label{tab:obs}}

\tablehead{Epoch & Telescope & Date & ObsID & Exp. Time}

\startdata
1 & XMM & 2006 Jan 28 & 0312190401 & 8.8 \\
2 & Suzaku & 2007 Jan 23 & 701013010 & 100 \\
3 & Suzaku & 2007 Jul 21 & 702079010 & 40  \\
4 & NuSTAR & 2017 Oct 14 & 60368001002 & 21  \\
5 & NuSTAR & 2024 Jul 13 & 61002024002 & 21  \\
 & +XMM & - & 0942010101 & 20  \\
6 & NuSTAR & 2024 Jul 26 & 61002024004 & 18 \\
 & +XMM & - & 0942010201 & 21  \\
7 & NuSTAR & 2024 Aug 23 & 61002024006 & 21  \\
 & +XMM & - & 0942010301 & 12  \\
8 & NuSTAR & 2024 Oct 23 & 61002024008 & 20  \\
9 & NuSTAR & 2025 Feb 5 & 61002024010 & 25  \\
 & +XMM & - & 0942010401 & 18  \\
\hline
\enddata
\tablecomments{Exp. time is the effective exposure time given in kiloseconds.}
\end{deluxetable}

\subsection{XMM-Newton}


NGC\,1142 was observed by XMM-Newton 5 times with the first observation being on January 28, 2006 and the other four in AO-23 (ID:094201, P.I. Pizzetti), accompanying the NuSTAR observations in the Cycle 10 monitoring campaign. All observations were reduced according to the standard procedures using SAS \citep{jansen_xmm}. All event files were cleaned for flares and the cleaned exposure time used for spectral extraction is shown in Table \ref{tab:obs}. We use only the data obtained from the EPIC PN camera because the quality of the MOS1 and MOS2 data were much lower and increased the computation time of the analysis while not adding any additional information. We used circular extraction regions for the source and background spectra. The background region was chosen as close to the Y position of the source on the same detector as possible, while avoiding contamination from other X-ray emission. We used a radius of 45" for the background region while the source region was chosen to optimize the signal to noise ratio (SNR) of the data following \cite{Piconcelli2004}. The extraction radii for the source regions ranged from $\sim$19" to $\sim$36" while the SNR ranged from $\sim$28 to $\sim$45. The spectra were binned using the \texttt{ftgrouppha} routine with the optimal binning scheme from \cite{Kaastra_opt16} and the data were fit using the Cash statistic \citep{cstat79}. 

\subsection{NuSTAR}

NGC\,1142 was observed by NuSTAR 6 times with the first observation being on October 14, 2017 and the other five in Cycle 10 (ID:00010204, P.I. Torres-Alb\`a). All observations were reduced following standard procedures using \texttt{nupipeline} and \texttt{nuproducts} scripts. The source and background spectral extraction regions were chosen similarly to the XMM-Newton data. In this case, we used a background radius of 100" and the source radius ranged from $\sim$55" to $\sim$70" while the SNR ranged from $\sim$25 to $\sim$32. We used the data from both the FPMA and FPMB modules. These spectra were also binned using the optimal binning scheme in \texttt{ftgrouppha}.

\subsection{Suzaku}

NGC\,1142 was observed by Suzaku twice in 2007, separated by 6 months. Both observations were reduced following the standard procedure outlined in the Suzaku Data Reduction Guide\footnote{https://heasarc.gsfc.nasa.gov/docs/suzaku/analysis/abc/}. The \texttt{aepipeline} script was used to reprocess the data and \texttt{xisrmfgen}, \texttt{xissimarfgen}, and \texttt{xisnxbgen} were used to generate the response, ancillary, and non X-ray background files, respectively. The source extraction radius was 250". The data from the two frontside illuminated chips (XI0 and XI3) were combined into one spectrum and fit simultaneously with the spectrum from the backside illuminated chip XI1. Due to the large number of counts in these observations, the spectra were grouped to 50 counts per bin.

\section{Spectral Modeling} \label{sec:models}

We use three different torus models to account for the reflection component in the spectrum of NGC\,1142. For all three models, the primary X-ray emission is assumed to be a cutoff powerlaw. We assume a high-energy cutoff of 300\,keV \citep[e.g.][]{Balokovic20} and leave this parameter frozen in all fits, since the data are not able to constrain it. We also find that the soft emission ($\lesssim2$\,keV) is best fit with two \texttt{apec} components. Below, we provide the exact model definitions used along with a brief description for each of the three models.

We use the \texttt{Xspec} software version 12.15.0 \citep{Arnaud96}. We perform simultaneous spectral fitting across all 9 observations in which the $N_H$ and cross-normalizations $(C_i)$ are allowed to vary between observations\footnote{Cross-normalization between simultaneous NuSTAR and XMM-Newton observations was left fixed at 1 to reduce the number of free parameters and to avoid unphysical fits in which the NuSTAR data preferred a reflection dominated fit while the simultaneous XMM-Newton data preferred a less obscured solution.} but the parameters describing the reprocessed emission are tied across all epochs \citep[see our other works][for more details]{marchesi_compton-thick_2022,pizzetti_multi-epoch_2022,NTA23,Pizzetti2025,NTA25,Sengupta25}. In addition to this fitting procedure, we also performed fits in which the photon index of the main powerlaw (which in previous works was tied between epochs) was allowed to vary between observations. Finally, we also attempted fits on each epoch individually. We compare the results of these procedures in Section \ref{sec:model_comparisons}.

We begin the fitting process by fitting only the primary emission normalization, while assuming a photon index of $\Gamma=1.8$ and a line-of-sight column density of $N_H=10^{24}$\,cm$^{-2}$ which are intermediate values for previously obtained results on observations 1--3 \citep[see Table 12 in][]{Pizzetti2025}. Once the normalization is found assuming these `reasonable' values, we perform a fit with only $\Gamma$, $N_{H,i}$, $C_i$, and the \texttt{apec} temperatures $(kT_{a,b})$ and normalizations left free. The geometrical torus parameters were left fixed to the values indicated in the subsequent sections for the initial fit. We then thawed each of the three torus parameters one at a time and refit. We then permuted the order in which the torus parameters were thawed to ensure we reached the same fit. The \texttt{error} routine was performed for each parameter until a stable fit was obtained. Further parameter space exploration was performed by using the \texttt{steppar} routine over many parameter combinations and ranges. 

Once we were confident the global minimum had been found, we constructed an MCMC chain using the Goodman-Weare algorithm with 100 walkers and a chain length of $10^7$ steps. A visual inspection of the chains indicated that a burn in phase of $10^3-10^4$ was enough to ensure they all reached a stationary distribution, but to be conservative we discard the first $10^5$ steps of all of our parameter posterior distributions.

\subsection{UXCLUMPY}

\texttt{UXCLUMPY} \citep{buchner_x-ray_2019} has been widely used to model AGN spectra \citep[e.g.][]{Silver22b,Lefkir2023,Kayal2023,NTA23,Belvedersky2025,Boorman2025,Cox25,Esparza-Arredondo2025}. It assumes a distribution of individual clouds with various column densities and sizes. Also included is a ring of Compton-thick $(N_H>1.5\times10^{24}$\,cm$^{-2}$) material that can account for the excess reflection in some sources \citep[e.g.][]{NTA23,Pizzetti2025}. The parameters defining the geometrical properties are \texttt{TORsigma}, \texttt{CTKcover}, and \texttt{Theta\_inc}. \texttt{TORsigma} $(\sigma_{\text{Tor}})$ defines the width of the normally distributed cloud locations away from the equatorial and it is varied between $6^{\circ}$ and $84^{\circ}$. \texttt{CTKcover} $(f_{\rm CT})$ defines the covering fraction of a ring of Compton-thick clouds $(\log(N_H)\sim25)$ and is varied between 0 and 0.6. Finally, \texttt{Theta\_inc} $(\theta_{\rm inc})$ is the inclination angle measured from the polar direction and is varied between $0^{\circ}$ (face-on) and $90^{\circ}$ (edge-on). 

Emission can also be scattered off of ionized gas into the line of sight allowing an essentially unobscured primary component to be visible in otherwise obscured directions. This is implemented in \texttt{UXCLUMPY} with the \texttt{uxclumpy-omni.fits} table. We link all the parameters of this component to those of the \texttt{uxclumpy.fits} table but multiply it by a constant (the scattering fraction, $f_s$) which we allow to vary between 0 and 0.1.

The precise model definition in \texttt{Xspec} is
\begin{align*}
	\texttt{UXCLUMPY}  = &\text{TBabs}* \\
	& \big( \text{constant}*(\text{atable\{uxclumpy.fits\}} \\
	& + \text{constant}*\text{atable\{uxclumpy-omni.fits\}}) \\
	& + \text{apec} + \text{apec} \big)
\end{align*}
where the first constant is the cross-normalization $(C_i)$ to account for flux variability in the primary coronal emission between observations and the second constant is $f_s$. TBabs accounts for Galactic absorption and is fixed at $5.56\times10^{20}$\,cm$^{-2}$ \citep{Kalberla2005}.

\subsection{XSKIRTor}

\texttt{XSKIRTor} \citep[][Vander Meulen et al. submitted]{Meulen23} is a new model that our group is beginning to use \citep[e.g.][Marchesi et al. in prep.]{Gianolli2026}. \texttt{XSKIRTor} has the same geometry as \texttt{borus02} \citep{balokovic_new_2018} which our previous works have used. However, \texttt{XSKIRTor} includes a more complete physical treatment of the radiative transfer calculations. It is based off of the \texttt{SKIRT} Monte Carlo Radiative transfer code \citep{Baes03,Baes11,Camps15,Camps20} which can handle complex, 3D geometries while incorporating scattering on free and bound electrons, absorption and reflection by cold gas and dust. The \texttt{XSKIRTor} tables assume a homogeneous sphere with conical cutouts at the poles. The parameters defining the geometrical properties are \texttt{CF}, \texttt{logNH}, and \texttt{cosi}. \texttt{CF} is the covering factor of the sphere defined by the opening angle of the polar cutouts and it is varied between 0.25 and 0.85. \texttt{logNH} is the $\log_{10}$ of the average column density of the material and is varied between 22 and 25. \texttt{cosi} is the cosine of the inclination angle measured from the polar direction and is varied between 0 (edge-on) and 1 (face-on).

The absorbed primary emission is added to the reprocessed emission separately by multiplying the \texttt{zcutoffpl} model in \texttt{Xspec} with the \texttt{XSKIRTor} extinction table. The line-of-sight $N_H$ and the average $N_H$ are not tied together to simulate an inhomogeneous obscurer/reflector. The precise model definition in \texttt{Xspec} is
\small
\begin{align*}
	\texttt{XSKIRTor} = &\text{TBabs}* \\
	& \big( \text{constant}*(\text{atable\{xskirtor\_reprocessed.mod\}} \\
	& + \text{etable\{xskirtor\_extinction.mod\}}*\text{zcutoffpl} \\
	& + \text{constant}*\text{zcutoffpl}) + \text{apec} + \text{apec} \big)
\end{align*}\normalsize
where the cross-normalization and scattering fraction constants are the same as in the \texttt{UXCLUMPY} model setup. The parameters (photon index, normalization, high-energy cutoff and redshift) in the \texttt{zcutoffpl} models are tied to the parameters in the xskirtor\_reprocessed.mod component.

\subsection{RXTorusD}

\texttt{RXTorusD} \citep{Ricci_Paltani23} is another new model widely used since its release \citep[e.g.][Marchesi et al. in prep]{Ricci23b,Peca2025,Gianolli2026,Silver2026}. \texttt{RXTorusD} has the same geometry as the \texttt{MYTorus} model \citep{murphy_x-ray_2009} but allows for a variable covering factor. It is calculated using the \texttt{REFLEX} radiative transfer code \citep{Paltani17} and extends the \texttt{RXTorus} model presented there to include dust scattering. The parameters defining the geometrical properties are \texttt{r/R}, \texttt{NHeq}, and \texttt{Inclination}. The \texttt{r/R} parameter is the ratio between the radius of the torus to the distance from the X-ray source and defines the covering factor. It is varied between 0 and 1 with 0.5 corresponding to an opening angle of $60^{\circ}$ which is equivalent to the \texttt{MYTorus} geometry. \texttt{NHeq} is the column density of the torus in the equatorial plane (i.e., integrated over the full diameter of the torus).

We include the absorbed primary component and elastic scattered emission in a similar manner to the \texttt{XSKIRTor} setup. Again, the line-of-sight column density is not tied to the column density of the reflector in any way. The precise model definition in \texttt{Xspec} is
\small
\begin{align*}
	\texttt{RXtorusD} = &\text{TBabs}* \\
	& \big( \text{constant}*(\text{atable\{RXTorus\_1.0\_1.0\_rprc.mod\}} \\
	& + \text{etable\{RXTorus\_1.0\_1.0\_cont.mod\}}*\text{zcutoffpl} \\
	& + \text{constant}*\text{zcutoffpl}) + \text{apec} + \text{apec} \big)
\end{align*}\normalsize
where everything is set up as in the \texttt{XSKIRTor} model.

\section{Results} \label{sec:results}


\begin{deluxetable}{c|ccc}[ht!]

\tablecaption{NGC\,1142 best-fit results. \label{tab:results_constant_gamma}}

\tablehead{Model &  UXCLUMPY &  XSKIRTor & RXTorus  }

\startdata
cstat/d.o.f & 2370/2027 & 2379/2027 & 2378/2027 \\
\hline
$\Gamma$ & $2.06^{+0.04}_{-0.12}$ & $1.64^{+0.05}_{-0.08}$ & $2.06^{+0.10}_{-0.07}$ \\ 
$\sigma_{\rm Tor}$ & $42^{+21}_{-14}$ & -- & -- \\ 
CTKcover & $0.00^{+0.23}_{-p}$ & -- & -- \\ 
$\theta_{\rm inc}$ & $60^{+19}_{-39}$ & $79^{+u}_{-2.7}$ & $2.6^{+1.3}_{-0.81}$ \\ 
$C_f$ & -- & $0.42^{+0.07}_{-0.05}$ & -- \\ 
$\log(N_{H,avg})$ & -- & $23.5^{+0.06}_{-0.04}$ & -- \\ 
$N_{H,eq}$ & -- & -- & $207^{+30}_{-23}$ \\ 
$r/R$ & -- & -- & $1^{+p}_{-0.01}$ \\ 
norm $(10^{-3})$ & $8.8^{+1.3}_{-2.8}$ & $2.9^{+0.8}_{-0.7}$ & $5.2^{+1.8}_{-1.1}$ \\ 
$f_s$  $(10^{-2})$ & $1.2^{+0.5}_{-0.2}$ & $0.6^{+0.2}_{-0.1}$ & $0.34^{+0.1}_{-0.1}$ \\ 
$kT_a$ & $0.84^{+0.08}_{-0.05}$ & $0.82^{+0.06}_{-0.06}$ & $0.84^{+0.06}_{-0.06}$ \\ 
norm$_a$ $(10^{-5})$ & $1.4^{+0.2}_{-0.2}$ & $1.5^{+0.2}_{-0.2}$ & $1.4^{+0.2}_{-0.2}$ \\ 
$kT_b$ & $0.18^{+0.09}_{-0.04}$ & $0.16^{+0.07}_{-0.03}$ & $0.17^{+0.10}_{-0.03}$ \\ 
norm$_b$ $(10^{-5})$ & $1.6^{+2.2}_{-0.3}$ & $1.8^{+3.1}_{-0.3}$ & $1.4^{+2.0}_{-0.3}$ \\ 
\hline 
$C_1$ & $1.61^{+0.20}_{-0.20}$ & $1.52^{+0.22}_{-0.17}$ & $1.75^{+0.26}_{-0.21}$ \\ 
$C_2$ & $2.69^{+0.29}_{-0.24}$ & $2.76^{+0.33}_{-0.24}$ & $2.96^{+0.35}_{-0.33}$ \\ 
$C_3$ & $2.78^{+0.29}_{-0.30}$ & $2.84^{+0.32}_{-0.29}$ & $2.94^{+0.36}_{-0.36}$ \\ 
$C_4$ & $2.02^{+0.42}_{-0.30}$ & $1.64^{+0.25}_{-0.24}$ & $1.70^{+0.24}_{-0.24}$ \\ 
$C_5$ & $1^*$ & $1^*$ & $1^*$ \\ 
$C_6$ & $0.92^{+0.14}_{-0.10}$ & $0.90^{+0.14}_{-0.11}$ & $0.94^{+0.14}_{-0.12}$ \\ 
$C_7$ & $1.14^{+0.18}_{-0.10}$ & $1.09^{+0.15}_{-0.13}$ & $1.13^{+0.19}_{-0.12}$ \\ 
$C_8$ & $1.22^{+0.35}_{-0.19}$ & $1.21^{+0.28}_{-0.23}$ & $1.18^{+0.48}_{-0.18}$ \\ 
$C_9$ & $0.83^{+0.11}_{-0.10}$ & $0.79^{+0.13}_{-0.09}$ & $0.80^{+0.13}_{-0.09}$ \\ 
\hline 
$N_{H,1}$ & $55^{+2}_{-4}$ & $52^{+5}_{-5}$ & $59^{+5}_{-4}$ \\ 
$N_{H,2}$ & $68^{+2}_{-4}$ & $69^{+5}_{-6}$ & $73^{+4}_{-3}$ \\ 
$N_{H,3}$ & $94^{+4}_{-6}$ & $97^{+7}_{-8}$ & $104^{+7}_{-8}$ \\ 
$N_{H,4}$ & $199^{+50}_{-30}$ & $156^{+18}_{-17}$ & $268^{+162}_{-56}$ \\ 
$N_{H,5}$ & $123^{+14}_{-13}$ & $126^{+15}_{-10}$ & $145^{+45}_{-19}$ \\ 
$N_{H,6}$ & $117^{+13}_{-11}$ & $119^{+12}_{-12}$ & $136^{+28}_{-14}$ \\ 
$N_{H,7}$ & $129^{+14}_{-13}$ & $125^{+11}_{-12}$ & $157^{+33}_{-19}$ \\ 
$N_{H,8}$ & $137^{+38}_{-23}$ & $134^{+27}_{-20}$ & $189^{+1571}_{-58}$ \\ 
$N_{H,9}$ & $102^{+8}_{-9}$ & $101^{+13}_{-10}$ & $109^{+15}_{-9}$ \\ 
\hline
\enddata
\tablecomments{$N_{H,eq}$ and $N_{H,i}$ have units of $10^{22}$\,cm$^{-2}$. A superscript of $u$ indicates that the parameter is unconstrained in that direction, while a superscript of $p$ indicates that the most probable value for the parameter is pegged at that limit. A superscript of $*$ indicates the parameter was frozen at that value.}
\end{deluxetable}


\begin{table}[t]
\centering
\caption{Fraction of epoch combinations resulting in observed variability. Values in parentheses are calculated from the values in Table \ref{tab:results} which allow photon index variability.}\label{tab:combos}

\begin{tabular}{c|ccc}
\hline
\hline

$N_{\rm Obs}$ & $f_{\rm UXCLUMPY}$ &  $f_{\rm XSKIRTor}$ & $f_{\rm RXTorus}$ \\
\hline
 \multicolumn{4}{c}{\cite{NTA23} method}  \\
\hline

9 & 1.0 (1.0) & 1.0 (1.0) & 1.0 (1.0) \\
8 & 1.0 (1.0) & 1.0 (1.0) & 1.0 (1.0) \\
7 & 1.0 (1.0) & 1.0 (1.0) & 1.0 (1.0) \\
6 & 1.0 (0.99) & 0.98 (0.99) & 1.0 (1.0) \\
5 & 0.98 (0.95) & 0.91 (0.95) & 0.98 (0.98) \\
4 & 0.93 (0.88) & 0.80 (0.88) & 0.92 (0.93) \\
3 & 0.79 (0.80) & 0.65 (0.74) & 0.76 (0.81) \\
2 & 0.61 (0.53) & 0.42 (0.44) & 0.53 (0.56) \\
\hline
\multicolumn{4}{c}{\cite{Nowack16} method}  \\
 \hline
 9 & 1.0 (1.0) & 1.0 (1.0) & 1.0 (1.0) \\
8 & 1.0 (1.0) & 1.0 (1.0) & 1.0 (1.0) \\
7 & 1.0 (1.0) & 0.97 (1.0) & 1.0 (1.0) \\
6 & 0.99 (0.99) & 0.93 (0.99) & 1.0 (0.99) \\
5 & 0.94 (0.95) & 0.87 (0.95) & 0.98 (0.95) \\
4 & 0.88 (0.87) & 0.78 (0.88) & 0.90 (0.88) \\
3 & 0.73 (0.74) & 0.64 (0.74) & 0.77 (0.76) \\
2 & 0.47 (0.47) & 0.42 (0.44) & 0.50 (0.53) \\
\hline
\multicolumn{4}{c}{Difference distributions (99\,\% C.I.)}  \\
\hline
2 & 0.75(0.61) & 0.67(0.67) & 0.72(0.67) \\
\hline
\end{tabular}
\end{table}


\begin{deluxetable}{c|ccc}[ht!]

\tablecaption{Best-fit results with variable photon index. \label{tab:results}}

\tablehead{Model & UXCLUMPY &  XSKIRTor & RXTorus  }

\startdata
cstat/d.o.f & 2257/2019 & 2225/2019 & 2269/2019 \\
\hline
$\sigma_{\rm Tor}$ & $35^{+21}_{-6.6}$ & -- & -- \\ 
CTKcover & $0.00^{+0.11}_{-p}$ & -- & -- \\ 
$\theta_{\rm inc}$ & $62^{+16}_{-33}$ & $70^{+2.9}_{-3.4}$ & $2.3^{+7.6}_{-0.45}$ \\ 
$C_f$ & -- & $0.45^{+0.06}_{-0.06}$ & -- \\ 
$\log(N_{H,avg})$ & -- & $23.7^{+0.1}_{-0.07}$ & -- \\ 
$N_{H,eq}$ & -- & -- & $158^{+21}_{-18}$ \\ 
$r/R$ & -- & -- & $0.45^{+0.06}_{-0.08}$ \\ 
norm $(10^{-3})$ & $8.8^{+2.5}_{-2.1}$ & $3.0^{+0.9}_{-0.6}$ & $4.4^{+1.2}_{-0.9}$ \\ 
$f_s$ $(10^{-2})$ & $0.82^{+0.33}_{-0.17}$ & $0.47^{+0.11}_{-0.10}$ & $0.21^{+0.05}_{-0.05}$ \\ 
$kT_1$ & $0.88^{+0.07}_{-0.06}$ & $0.84^{+0.07}_{-0.05}$ & $0.93^{+0.05}_{-0.08}$ \\ 
norm$_1$ $(10^{-5})$ & $1.5^{+0.17}_{-0.18}$ & $1.5^{+0.13}_{-0.20}$ & $1.6^{+0.16}_{-0.19}$ \\ 
$kT_2$ & $0.20^{+0.08}_{-0.04}$ & $0.18^{+0.07}_{-0.02}$ & $0.25^{+0.03}_{-0.07}$ \\ 
norm$_2$ $(10^{-5})$ & $1.7^{+1.2}_{-0.30}$ & $1.8^{+1.8}_{-0.24}$ & $1.9^{+0.78}_{-0.24}$ \\ 
\hline 
$C_1$ & $1.11^{+0.32}_{-0.22}$ & $1.23^{+0.34}_{-0.25}$ & $1.33^{+0.45}_{-0.31}$ \\ 
$C_2$ & $3.56^{+0.60}_{-0.53}$ & $3.85^{+0.75}_{-0.50}$ & $4.40^{+1.09}_{-0.65}$ \\ 
$C_3$ & $3.82^{+0.83}_{-0.51}$ & $4.25^{+0.82}_{-0.62}$ & $4.89^{+1.24}_{-0.82}$ \\ 
$C_4$ & $1.42^{+0.65}_{-0.42}$ & $0.88^{+0.30}_{-0.19}$ & $0.90^{+0.35}_{-0.29}$ \\ 
$C_5$ & $1^*$ & $1^*$ & $1^*$ \\ 
$C_6$ & $0.89^{+0.21}_{-0.15}$ & $0.89^{+0.20}_{-0.15}$ & $0.91^{+0.25}_{-0.20}$ \\ 
$C_7$ & $0.95^{+0.25}_{-0.19}$ & $0.88^{+0.22}_{-0.15}$ & $0.95^{+0.28}_{-0.18}$ \\ 
$C_8$ & $0.95^{+0.96}_{-0.43}$ & $0.89^{+0.44}_{-0.37}$ & $1.10^{+0.55}_{-0.33}$ \\ 
$C_9$ & $0.86^{+0.20}_{-0.17}$ & $0.90^{+0.23}_{-0.16}$ & $0.93^{+0.29}_{-0.18}$ \\ 
\hline 
$\Gamma_1$ & $1.81^{+0.13}_{-0.15}$ & $1.50^{+0.12}_{-0.13}$ & $1.65^{+0.13}_{-0.13}$ \\ 
$\Gamma_2$ & $2.26^{+0.06}_{-0.14}$ & $1.93^{+0.09}_{-0.08}$ & $2.09^{+0.09}_{-0.06}$ \\ 
$\Gamma_3$ & $2.34^{+0.09}_{-0.12}$ & $2.01^{+0.11}_{-0.07}$ & $2.22^{+0.08}_{-0.10}$ \\ 
$\Gamma_4$ & $1.95^{+0.11}_{-0.10}$ & $1.43^{+0.10}_{-0.10}$ & $1.49^{+0.10}_{-0.22}$ \\ 
$\Gamma_5$ & $2.01^{+0.10}_{-0.06}$ & $1.61^{+0.10}_{-0.09}$ & $1.67^{+0.09}_{-0.15}$ \\ 
$\Gamma_6$ & $1.99^{+0.11}_{-0.07}$ & $1.61^{+0.09}_{-0.11}$ & $1.68^{+0.10}_{-0.17}$ \\ 
$\Gamma_7$ & $1.95^{+0.07}_{-0.09}$ & $1.55^{+0.08}_{-0.09}$ & $1.65^{+0.07}_{-0.10}$ \\ 
$\Gamma_8$ & $2.03^{+0.13}_{-0.21}$ & $1.60^{+0.11}_{-0.14}$ & $1.71^{+0.12}_{-0.11}$ \\ 
$\Gamma_9$ & $2.06^{+0.08}_{-0.09}$ & $1.71^{+0.08}_{-0.08}$ & $1.83^{+0.07}_{-0.08}$ \\ 
\hline 
$N_{H,1}$ & $61^{+6}_{-5}$ & $55^{+7}_{-5}$ & $66^{+6}_{-5}$ \\ 
$N_{H,2}$ & $69^{+2}_{-4}$ & $61^{+5}_{-3}$ & $74^{+3}_{-3}$ \\ 
$N_{H,3}$ & $86^{+7}_{-7}$ & $78^{+10}_{-6}$ & $93^{+6}_{-6}$ \\ 
$N_{H,4}$ & $200^{+69}_{-51}$ & $174^{+38}_{-31}$ & $241^{+153}_{-32}$ \\ 
$N_{H,5}$ & $136^{+16}_{-14}$ & $154^{+35}_{-21}$ & $216^{+77}_{-41}$ \\ 
$N_{H,6}$ & $130^{+16}_{-15}$ & $143^{+34}_{-20}$ & $178^{+90}_{-25}$ \\ 
$N_{H,7}$ & $143^{+22}_{-18}$ & $143^{+22}_{-17}$ & $186^{+34}_{-23}$ \\ 
$N_{H,8}$ & $138^{+44}_{-36}$ & $126^{+36}_{-28}$ & $183^{+47}_{-38}$ \\ 
$N_{H,9}$ & $107^{+12}_{-9}$ & $105^{+16}_{-10}$ & $128^{+16}_{-14}$ \\ 
\hline 
\enddata
\end{deluxetable}


\begin{figure*}[htbp]
   \centering
   \includegraphics[scale=.33, trim={0 0 0 0}, clip]{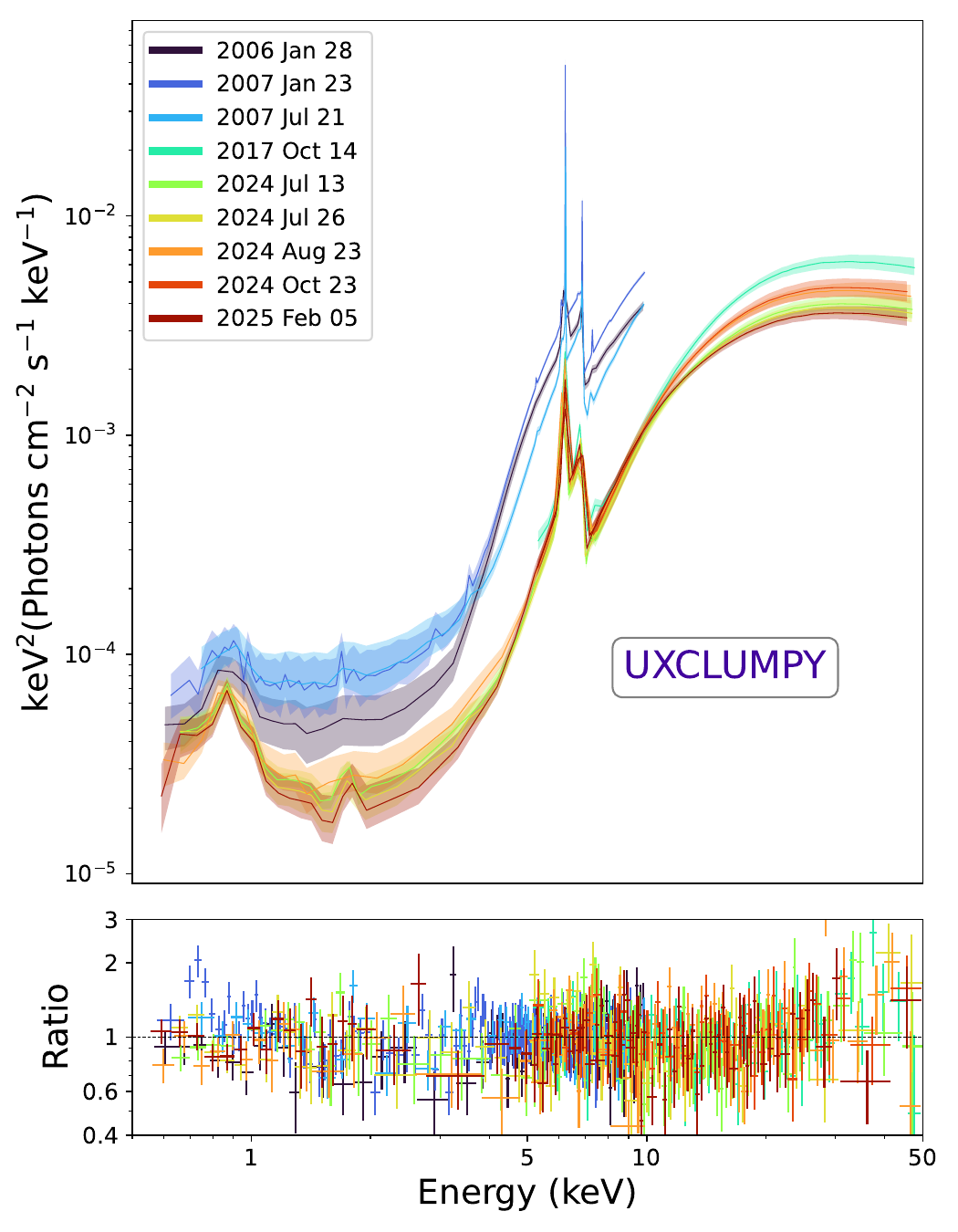}
   \includegraphics[scale=.33, trim={0 0 0 0}, clip]{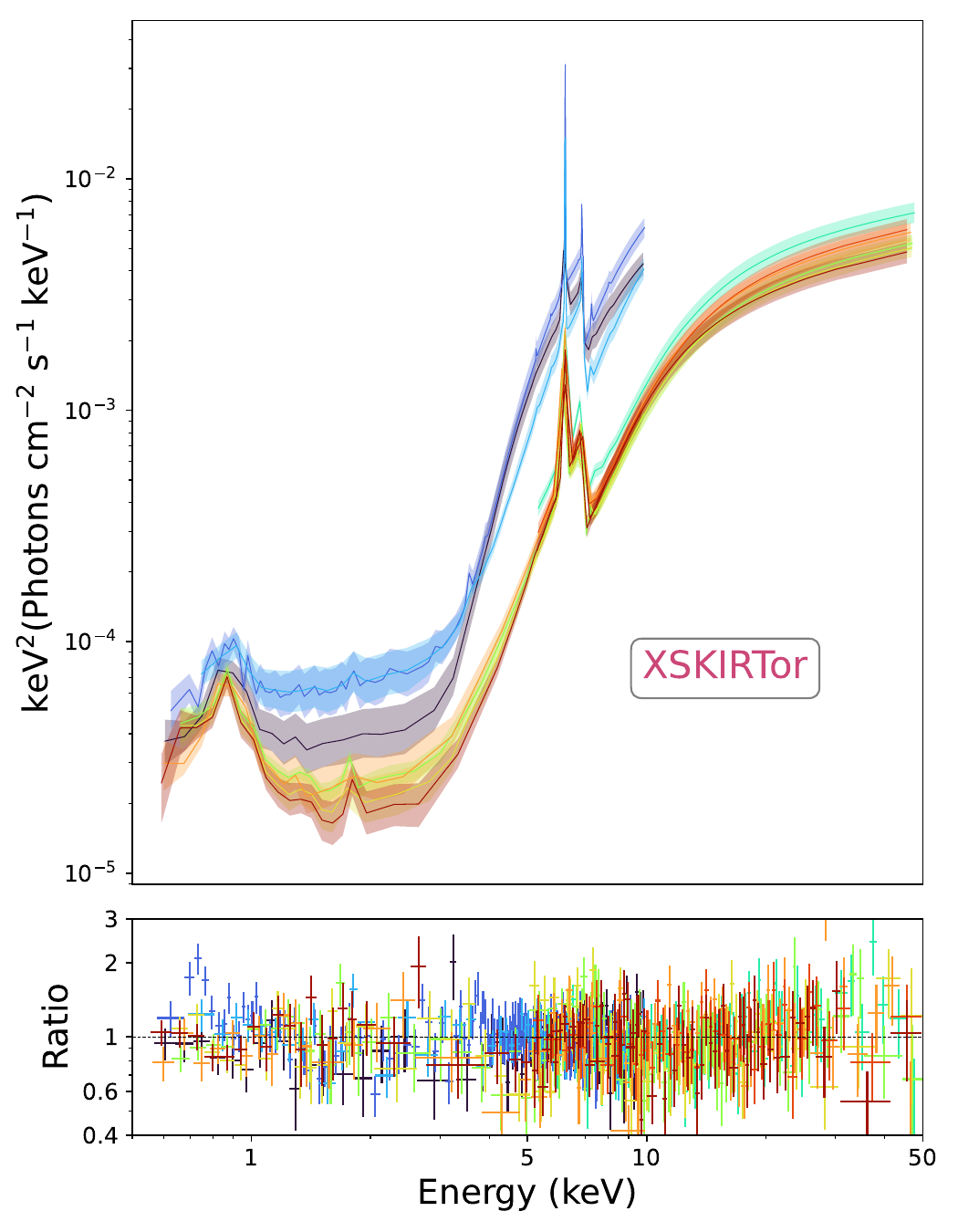}
   \includegraphics[scale=.33, trim={0 0 0 0}, clip]{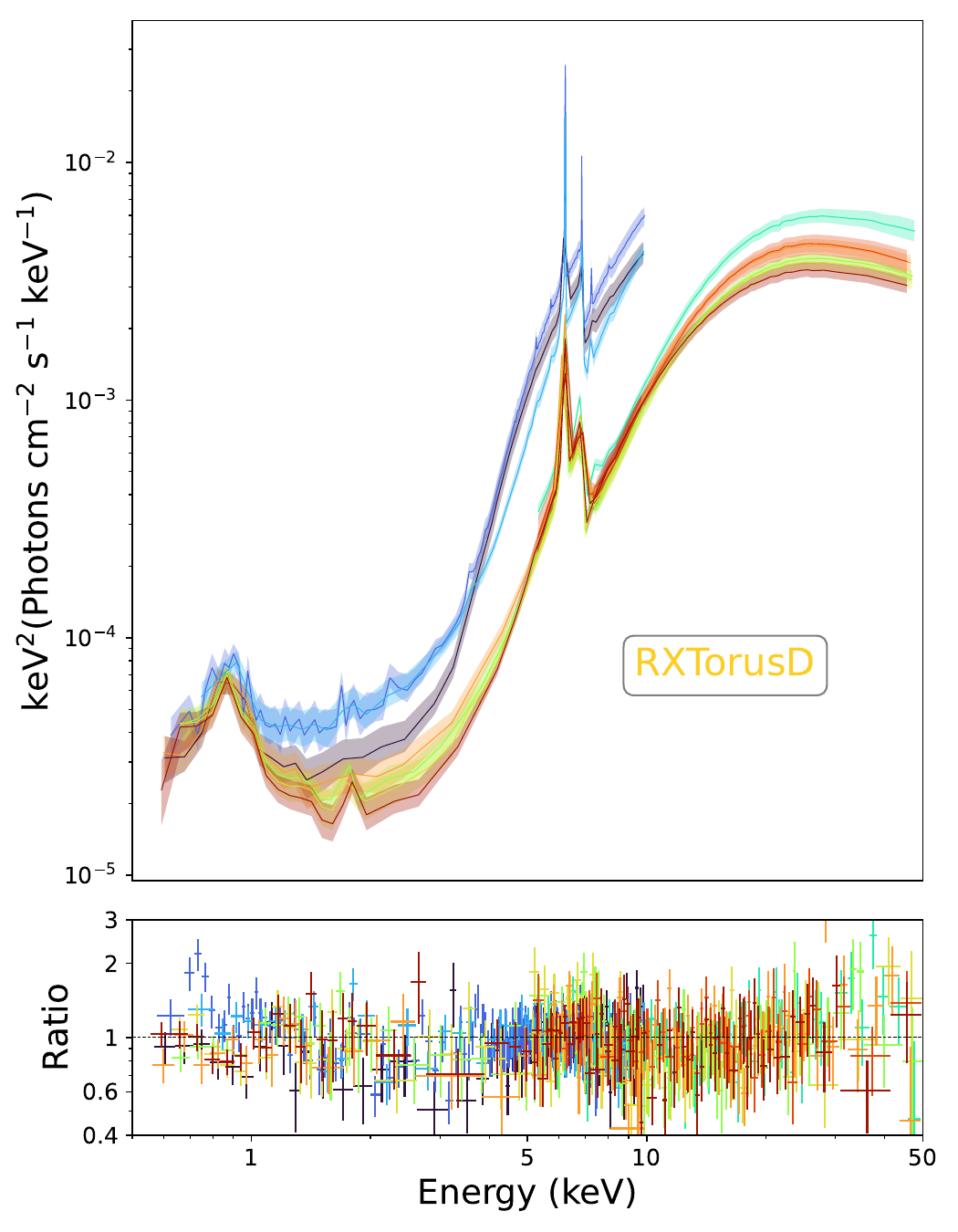}
   \caption{\textit{Left:} The best fit \texttt{UXCLUMPY} models for each of the 9 epochs with the residuals. The solid line indicates the best fit model while the shading indicates the 90\,\% credible interval from the MCMC chain. The plot has been rebinned to $3\sigma$ to improve clarity. \textit{Middle:} Same as left but for the \texttt{XSKIRTor} model. \textit{Right:} Same as left but for the \texttt{RXTorusD} model.}
   \label{fig:spectra}
\end{figure*}

The best-fit parameter values for all three models assuming a the same photon index in all epochs are shown in Table \ref{tab:results_constant_gamma}. The values shown are the mode of the posterior distribution, which is calculated using the Gaussian kernel density estimator available as \texttt{scipy.stats.gaussian\_kde} in the Python library SciPy \citep{scipy}. The lower and upper bounds provided for each parameter are calculated from the $5^{\rm th}$ and $95^{\rm th}$ percentiles of the posterior distribution, respectively. The best-fit spectra for each epoch along with the residuals for each of the three models are shown in Figure \ref{fig:spectra}.

We compute a $p$-value to indicate the probability of non-variability using equation 7 of \cite{NTA23} which is
\begin{equation}\label{eq:var_p_value}
	\chi^2 = \sum_{i=1}^{n}\frac{(N_{H,i}-\langle N_H\rangle)^2}{\delta^2N_{H,i}}
\end{equation}
where the $\delta N_{H,i}$ is chosen as the size of the error in the direction of the mean according to \cite{Barlow03}. We calculated the $p$-value for every possible combination of epoch subsets and use $p<0.01$ as the threshold for a variable classification. For the set of all 9 epochs, we find that the $p$-values for all three models are $\ll10^{-10}$, indicating very strong evidence of variability. Likewise, every possible combination of 8 and 7 observations yield $p\ll10^{-10}$ for all three models. When considering subcampaigns with fewer than 7 observations, the slight possibility of not detecting variability arises for the \texttt{XSKIRTor} model. The fractions of combinations that indicate variability for each subcampaign size are shown in Table \ref{tab:combos} under `\cite{NTA23} method' with the values in parentheses indicating the results allowing photon index to vary between observations. 

We also compute a $p$-value from the posteriors following the procedure outlined in \cite{Nowack16} and implemented, for example, in Appendix D of \cite{Zhao2024}. This procedure involves maximizing the likelihood of the posteriors assuming different $N_H$ values for each observation ($L_1^{\rm max}$) and assuming a single $N_H$ ($L_2^{\rm max}$). The $p$-value, interpreted as a `false alarm' probability, is then given as 
\begin{equation}
p=1-F\big(\chi^2=2(L_1^{\rm max}-L_2^{\rm max}), N_{\rm obs}-1\big)
\end{equation}
where $F$ is computed using \texttt{scipy.stats.chi2.cdf}. Again, we consider a campaign as truly variable if the false alarm probability is $p<0.01$. The fraction of subcampaigns that are variable are shown in Table \ref{tab:combos} under `\cite{Nowack16} method.' The trend shown in Table \ref{tab:combos} is consistent with that revealed in \cite{NTA25} and \cite{Cox26} which both show a probability of detecting variability to be  $\sim50$\,\% even for variable sources, when only 2 observations are available.

In addition, we determine variability of each individual pair of observations. We do this by creating a difference distribution $N_{H,i}-N_{H,j}$ from the $N_H$ posteriors, for all $9\choose2$ combinations of observations $i$ and $j$.\footnote{We only compute differences between epochs from the same walker on the same step of the chain, so that potential degeneracies are automatically taken care of.} If the central 99\,\% of the difference distribution does not include 0, then the observation pair is declared variable. The fraction of variable pairs determined in this way is shown in the bottom row of Table \ref{tab:combos}. These fractions can be compared to the $N_{\rm Obs}=2$ rows for the other two methods. This method seems to identify more observation pairs as variable compared to the $p$-value methods. This may be due to degeneracies between the observations. In the $p$-value methods, these degeneracies can only result in larger error bars. However, with the information contained in the posteriors, the difference distribution results in a more precise variability measurement since subtracting the posteriors of degenerate parameters can result in a narrower distribution than either of the individual posteriors. We believe this method is likely the most robust way to identify variability. However, to be consistent with previous publications, we choose to show the more conservative variability results using equation \ref{eq:var_p_value}. A computation of variability using this method will be reported for all the sources in our sample \citep{pizzetti_multi-epoch_2022,NTA23,Pizzetti2025,NTA25,Sengupta25} once $N_H$ posterior distributions have been obtained in a future work.

\begin{figure*}[htbp]
   \centering
   \includegraphics[scale=.50]{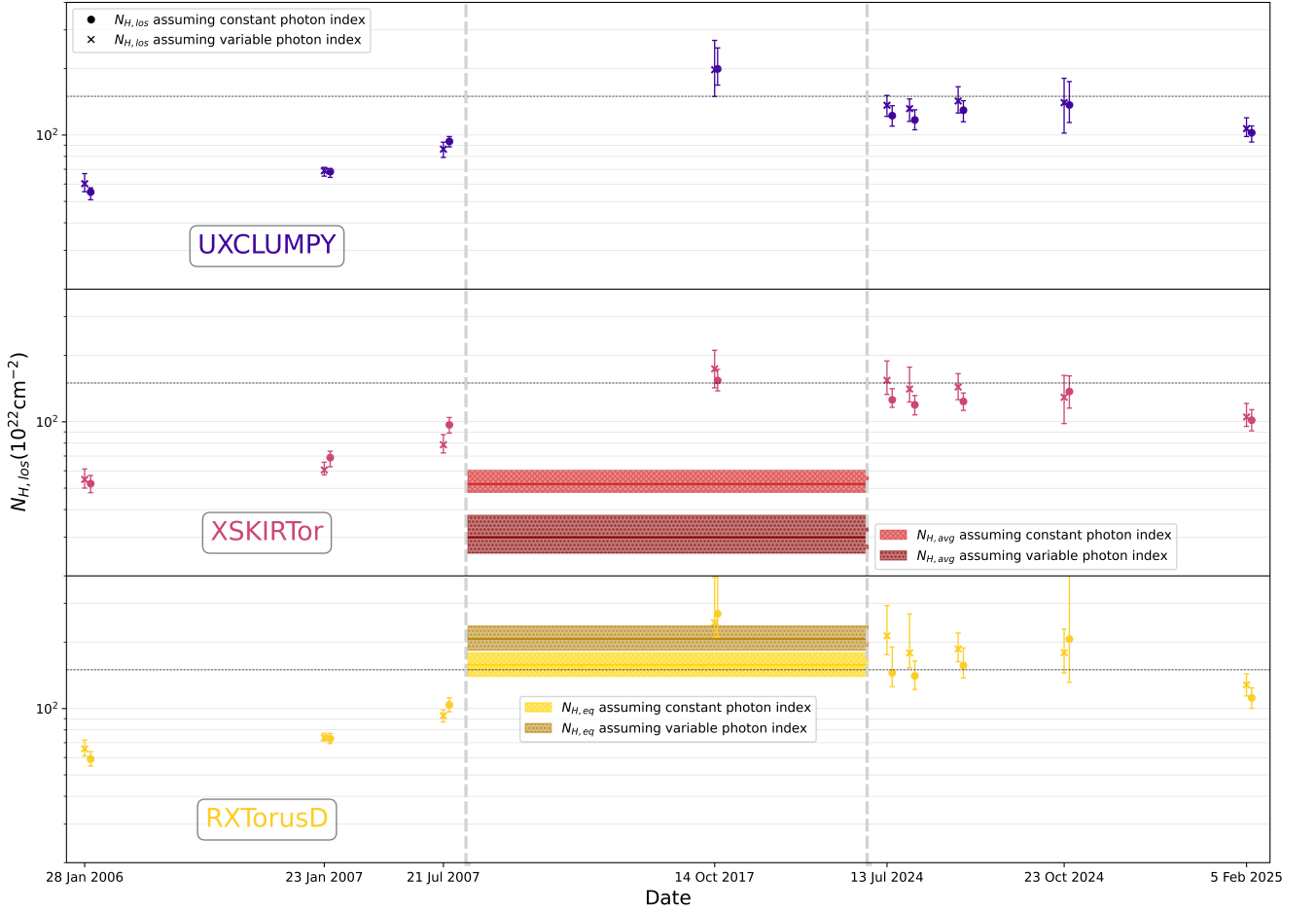}
   \caption{The line-of-sight $N_H$ as a function of time assuming a constant photon index (dots) and a variable photon index (crosses). The \textit{top} row shows the \texttt{UXCLUMPY} model results in blue while the \textit{middle} and \textit{bottom} rows show the \texttt{XSKIRTor} and \texttt{RXTorusD} results, respectively. The axes have the same scale for all three models. The $x$-axis has been chopped up (vertical, dashed lines) to get rid of dead time in the data. Note that the three time windows have different scales, with the earliest (left) window covering $\sim1.5$\,years while the most recent (right) window covers only $\sim7$\,months. The column densities responsible for the reflection component ($N_{H,avg}$ and $N_{H,eq}$ for \texttt{XSKIRTor} and \texttt{RXTorusD}, respectively) are shown as the horizontal bands in the middle time window. The values assuming a constant photon index are shown in the lighter shading with cross hatching and the values assuming a variable photon index are shown in the darker shading with dotted hatching. The Compton-thick threshold is shown as a black, dashed line at $N_{H,los}=150\times10^{22}$\,cm$^{-2}$.}
   \label{fig:nH_v_time}
\end{figure*}

Figure \ref{fig:nH_v_time} shows the line-of-sight $N_H$ as a function of time for all three models assuming both a constant (dots) and variable (crosses) photon index. Also shown are the column densities required to explain the reflection component. For the \texttt{XSKIRTor} model, this is average torus column density $N_{H,avg}$. For the \texttt{RXTorusD} model, this is column density of the torus in the equatorial plane $N_{H,eq}$. There is no good comparison to these column densities in the \texttt{UXCLUMPY} model due to the way the clumpy cloud distribution is defined. In the \texttt{XSKIRTor} model, $N_H\geq N_{H,avg}$ for all epochs which is consistent with the \texttt{borus02} results in \cite{Pizzetti2025}. However, $N_{H,eq}$ in \texttt{RXTorusD} is much higher than the $N_{H,eq}$ obtained with the \texttt{MYTorus} model in \cite{Pizzetti2025}. Based on the \texttt{RXTorusD} results, the line-of-sight $N_H$ varies over time having best-fit values that are both significantly lower and higher than $N_{H,eq}$, making this the only source (out of 39) studied thus far showing such behavior \citep[][]{pizzetti_multi-epoch_2022,NTA23,Pizzetti2025,NTA25,Sengupta25,Gianolli2026}.

We find the \texttt{UXCLUMPY} and \texttt{XSKIRTor} models prefer a more edge-on inclination, consistent with the results in \cite{Pizzetti2025}, who found a lower limit of $77^{\circ}$ with both the \texttt{UXCLUMPY} and \texttt{borus02} models. We also find that the Compton-thick inner ring component is not very significant, with an upper limit of \texttt{CTKcover}$<0.23$ which is consistent with the \cite{Pizzetti2025} measurement of \texttt{CTKcover}$<0.26$. When the photon index is allowed to vary between observations, the constraint improves further to \texttt{CTKcover}$<0.11$. The largest difference in the geometry parameters between the two fitting procedures is in the $r/R$ parameter in the \texttt{RXTorusD} model. If photon index is assumed constant, $r/R$ is pegged at its upper limit, whereas if photon index is allowed to vary, the parameter is well constrained to a more moderate value consistent with the covering factor found with \texttt{XSKIRTor}. However, given the face-on inclination, the line of sight does not intercept the torus in the case of a moderate covering factor. With the excellent agreement in inclination angle between the \texttt{UXCLUMPY} and \texttt{XSKIRTor} models, it is likely that the inclination angle measured by \texttt{RXTorusD} is not reliable.

We also performed spectral fits for each epoch separately, with nothing tied between observations. The primary difference is that every parameter is less constrained. However, they all remain consistent with the values obtained in the other two fitting procedures and importantly, none of the results on $N_H$ variability are significantly affected. For this reason, we will only consider the results in Tables \ref{tab:results_constant_gamma} and \ref{tab:results} for the discussion.

\begin{figure*}[htbp]
   \centering
   \includegraphics[scale=.58]{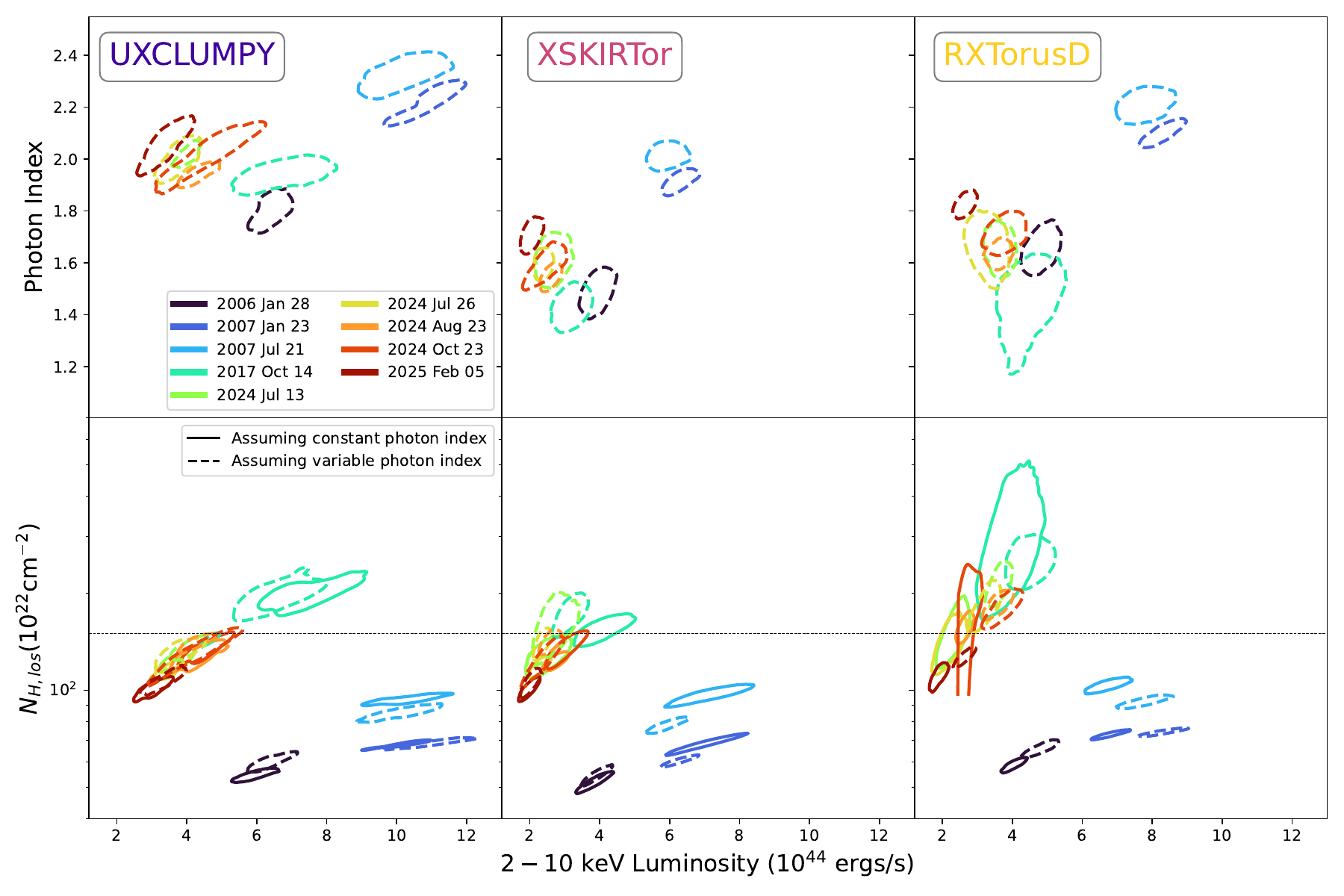}
   \caption{Contours enclosing the $90$\,\% credible intervals for $L_{\text{2--10\,keV}}$ with photon index (\textit{top row}) and $L_{\text{2--10\,keV}}$ with $N_{H,los}$ (\textit{bottom row}). The \textit{left} column shows the results for \texttt{UXCLUMPY} while the \textit{middle} and \textit{right} columns show the results for \texttt{XSKIRTor} and \texttt{RXTorusD}, respectively. The solid contours show the results assuming no variability in the photon index, while the dashed contours show the results allowing photon index to vary. The colors indicate the date of the observation and they are the same as in Figure \ref{fig:spectra}. The black, dashed line in the bottom row indicates the Compton thick threshold.}
   \label{fig:contours}
\end{figure*}

Figure \ref{fig:contours} shows the 90\,\% credible contours for the unabsorbed 2--10\,keV luminosity $(L_{\text{2--10\,keV}})$ with photon index and $N_H$. The luminosity posteriors were calculated from the photon index and normalization posteriors, assuming a cutoff powerlaw with $E_{\text{cut}}=300$. NGC\,1142 appears to have significantly increased in luminosity by a factor of $\sim$1.5--2 between the beginning of 2006 and the middle of 2007. This coincided with a significant increase in the column density. By 2017, the luminosity had decreased back to the 2006 level, however the column density reached its peak in our data, beyond the Compton-thick threshold according to the \texttt{UXCLUMPY} and \texttt{RXTorusD} models. Over the course of the 2024-2025 monitoring campaign, none of the parameters changed in such a significant manner, however, the source does appear to be dimming slightly, while getting softer. This is the opposite of the behavior in 2007 when the increasing luminosity was accompanied by a softer spectral shape. The same trend is seen in all three models.

\section{Discussion} \label{sec:discussion}

In this section, we compare the results between the different models and fitting methods. We discuss the observed variability dependence on the number of observations considered, as well as the timescale between the observations. We place these results in the context of other works.

\subsection{Model Comparisons}\label{sec:model_comparisons}

\begin{figure*}[htbp]
   \centering
   \includegraphics[scale=.58]{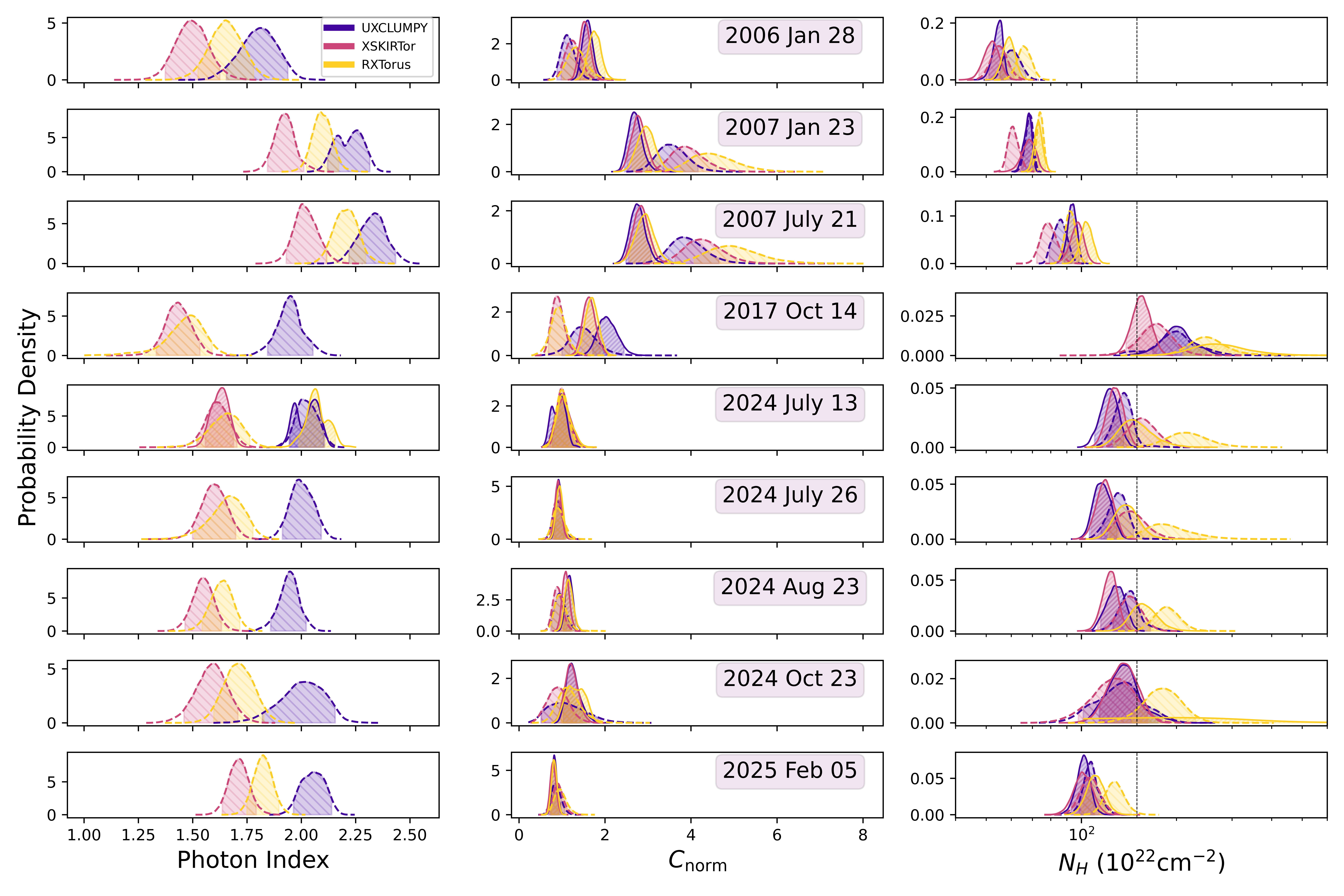}
   \caption{Posterior distributions of the photon index (left column), cross-normalization (middle column), and $N_H$ (right column) for each epoch (time increasing downward) and each model. The colors indicate the different models and are the same as Figure \ref{fig:nH_v_time}. The solid lines and shaded area with `/' hatching indicate the central $90$\,\% posterior mass for the fitting results assuming a constant photon index. The dashed lines with `\textbackslash' hatching show the same for the fits allowing photon index to vary between observations. The cross-normalizations are relative to the July 13, 2024 epoch, and the posteriors shown for that date have been divided by the `norm' values in Tables \ref{tab:results_constant_gamma} and \ref{tab:results} for the solid and dashed lines, respectively.}
   \label{fig:model_comparisons}
\end{figure*}

Fig \ref{fig:model_comparisons} shows the posterior distributions of photon index, cross-normalization, and $N_H$ for all 9 epochs. The three models are indicated by different colors while the posteriors assuming no variability in photon index are shown in the solid lines and the dashed lines indicate the posteriors where photon index was allowed to vary. From this plot, it is clear that the precise parameter values can differ between the models and fitting methods, however, the trends are in general agreement and the $N_H$ especially is largely consistent between the models as has been seen in many previous works \citep[e.g.][]{pizzetti_multi-epoch_2022,NTA23,Cox25,Pizzetti2025,NTA25,Sengupta25,Gianolli2026,Pal2026,Silver2026}. Importantly, the observed variability is not significantly affected (see the variable fractions in Table \ref{tab:combos}).

The photon index preferred by the \texttt{UXCLUMPY} model tends to be much softer than the preferred value by \texttt{XSKIRTor} for all epochs. \texttt{RXTorusD} tends to take a more intermediate value when variability is allowed, although it agrees with \texttt{UXCLUMPY} when the photon index is tied between observations. The most significant variability in the photon index comes in epochs 2 and 3 compared to the rest of the epochs. These epochs may be driving the \texttt{RXTorusD} photon index towards these higher values while the harder states may be driving the fits for the other models. However, it should also be noted that the covering factor ($r/R$) in the \texttt{RXTorusD} model is pegged at 1 when the photon index is tied. This is an unlikely physical geometry and is in stark disagreement with \texttt{XSKIRTor}. When the photon index is allowed to vary, the covering factor is well-constrained at a more reasonable value and is in perfect agreement with \texttt{XSKIRTor}. The large covering factor results in a much more prominent Compton reflection hump which could be pushing the photon index higher to compensate for the extra model flux at high energies. Indeed, when the no-variability fit is performed with $r/R$ fixed at 0.45, the photon index becomes much lower $(\sim1.8)$. This is in between the \texttt{UXCLUMPY} and \texttt{XSKIRTor} values, which is consistent with the general trend seen between the models in the variable fit, although the fit statistic is worse (cstat/d.o.f$=$2507/2028).

However, \texttt{RXTorusD} still prefers a more face-on inclination angle as opposed to the more edge-on solutions for \texttt{XSKIRTor} and \texttt{UXCLUMPY}. We also tried fitting the \texttt{RXTorusD} model with the inclination angle fixed at $60^{\circ}$. This fit also resulted in a photon index of $\sim1.8$ and a $r/R\sim0.45$. However, the fit statistic is again much worse (cstat/d.o.f$=$2604/2028). Further testing needs to be done on more sources to better understand these slight discrepancies in the models.

\subsection{Variability of Simulated Obscuration Curves}

\begin{figure*}[htbp]
   \centering
   \includegraphics[scale=0.58]{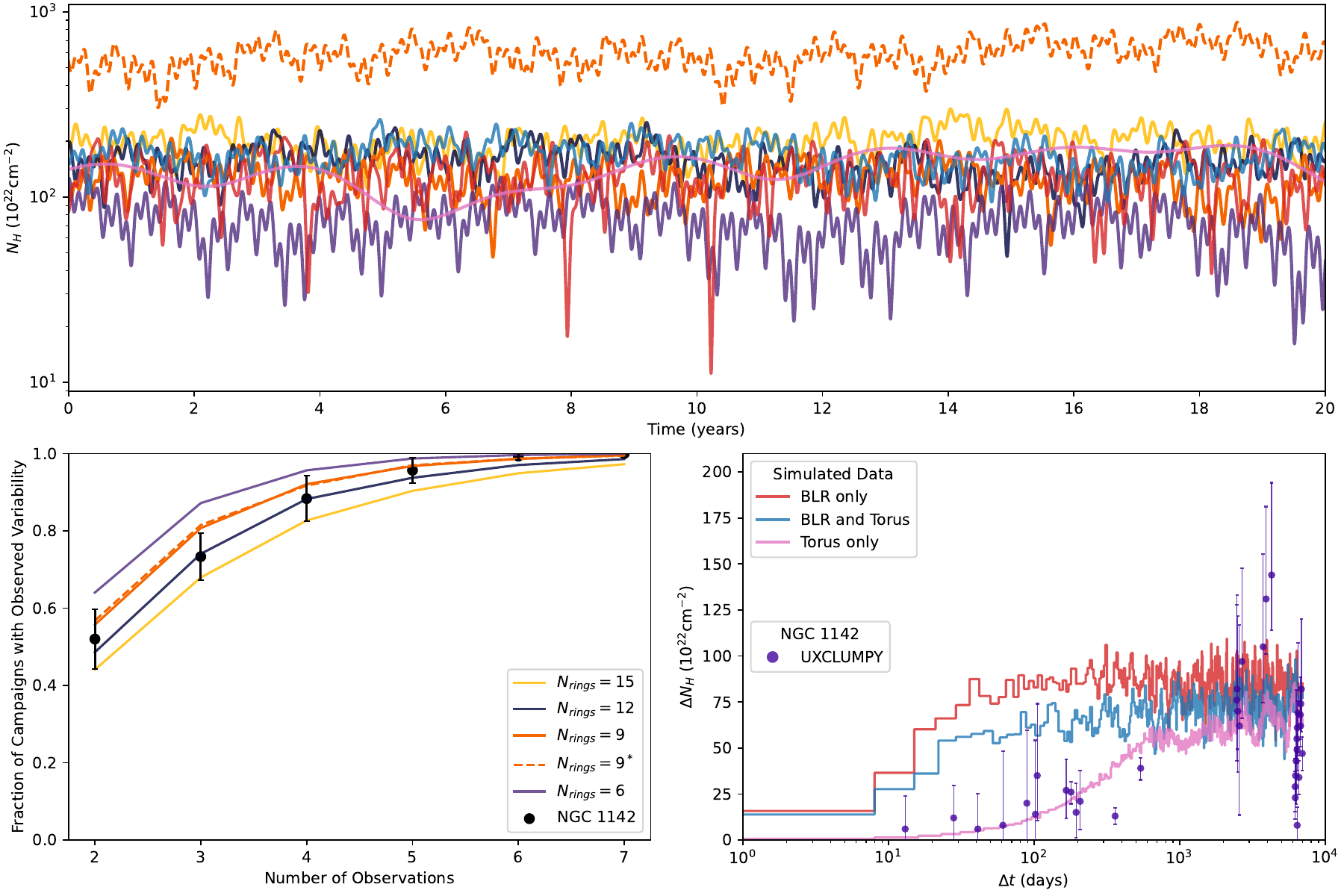}
   \caption{\textit{Top:} Examples of the simulated obscuration curves using Equation \ref{eq:cloud_sims}. The colors match the corresponding colors in the bottom two panels. \textit{Bottom Left:} The observed variable fraction of the curves as a function of the number of observations sampled. The colors represent the number of rings (or `clouds' participating in simultaneous absorption) simulated. The dashed, orange line represents the curve where the $N_H$ was artificially made higher. The black points are the average values from Table \ref{tab:combos} with errors calculated as the larger of the standard deviation between the models or the binomial errors added in quadrature. \textit{Bottom Right:} The magnitude of $N_H$ variability measured by the \texttt{UXCLUMPY} model as a function of the time between each pair of observations (purple dots). The values used come from Table \ref{tab:results_constant_gamma}. The upper errors are calculated by adding in quadrature the upper error of the higher $N_H$ value and the lower error value of the lower $N_H$ value. Similarly, the lower errors are calculated by adding in quadrature the lower error of the higher $N_H$ value and the higher error value of the lower $N_H$ value. The lines indicate the 90$^{\rm th}$ percentile of $\Delta N_H$ for simulated pairs considering three different scenarios: absorption only from broad line region scales (red), absorption from only torus scales (pink), and absorption from both simultaneously (blue). Equal frequency binning was performed with 200 points per bin.}
   \label{fig:cloud_sims}
\end{figure*}

The $N_H$ of NGC\,1142 clearly varies with time, as was originally shown over four observations in \cite{Pizzetti2025}. Significant variability was seen in $\sim$18\,months, with $N_H$ quickly increasing by a factor of almost 2. However, the recent monitoring campaign, which spans $\sim7$\,months, shows a much less dramatic change, and very little variability at all in most of the observations. 

A very strong dependence was identified in \cite{Cox26}, showing that the more observations a source has, the more likely spectral variability will be detected. Alternatively, the fewer observations a source has, the less likely variability will be detected, regardless of the true nature of the source. This trend is also present in other works \citep[e.g.][]{Laha2020,NTA23,NTA25,Gianolli2026}.

The trend remains present within the observation set of NGC\,1142 (see Table \ref{tab:combos}). When at least seven of the available observations are considered, all three models predict variability 100\,\% of the time. However, when only two observations are considered, NGC\,1142 would only present variability $\sim50$\,\% of the time. This underscores the necessity of long-term and diverse temporal sampling to observe variability on all timescales. Unfortunately, these data are not easy to obtain and high quality datasets are unlikely to be realized in the near future. 

To further explore the observational signatures and constraining power of our limited temporal sampling, we can generate qualitatively plausible obscuration curves, and sample observations from these. We construct these curves by considering concentric rings of azimuthally oscillating column density to approximate clouds at different orbital radii crossing in and out of the line of sight. Specifically, we use the function
\begin{equation}\label{eq:cloud_sims}
	N_{H,i}(t) = N_{H,\text{max}}\sin^2\left(\frac{\pi n}{T}t + \phi \right)
\end{equation}
to define the $N_H$ as a function of time, $t$, for the $i^{th}$ ring with a radius of $r$. Here, $n$ is the number of `clouds' that pass the line of sight per orbital time period, $T$, on a given ring. It is defined as $n=2\pi r f/D_{\text{cloud}}$ where $r$ is the orbital radius, $f$ is a volume filling factor, and $D_{\text{cloud}}$ is the diameter of the cloud. We assume a circular, Keplerian orbit so that $T=2\pi\sqrt{r^3/(GM)}$. $N_{H,max}$ is a decreasing function of radius with the clouds associated with the broad line region being $N_{H,max}\sim10^{23-24}$\,cm$^{-2}$ at $r\lesssim10^{-1}$\,pc, while the clouds farther out at $r\gtrsim1$\,pc have $N_{H,max}\sim10^{22-23}$\,cm$^{-2}$. The cloud diameter increases with radius from $D_{\text{cloud}}\sim10^{13-14}$\,cm at $r\lesssim10^{-1}$\,pc up to $D_{\text{cloud}}\sim10^{16-17}$\,cm at $r\gtrsim1$\,pc. We assume a volume filling factor of $f\sim0.01$ at all radii \citep[e.g.][]{Almeyda20,Matthews20}. Finally, $\phi$ is the phase, which we randomly generate for each ring.

This simple, semi-physical model is able to produce realistic time series of the line-of-sight $N_H$, with similar variability magnitudes and timescales that are seen in many sources. The top panel of Figure \ref{fig:cloud_sims} shows a few examples. While a detailed development and analysis of this model is beyond the scope of this work, we can test some of the basic predictions against the observables in our sample. To obtain the model predictions, we simulate 100 sources assuming a model with clouds located at 15, 12, 9, and 6 rings between $10^{-3}\text{\,pc}\leq r\leq10$\,pc. The number of rings can be thought of as the number of individual clouds that are simultaneously participating in an eclipse, and is indicated as $N_{\text{Clouds}}$ in Figure \ref{fig:cloud_sims}. The phase, $\phi$, is randomly chosen for each ring in each individual source. We then `observe' each of these sources 1000 times and compute the fraction of times we observe variability as a function of the number of observations considered. These are calculated in the same way as the fractions in Table \ref{tab:combos}. We assume an uncertainty of 5\,\% on these simulated observations. The dependencies are shown in the bottom left panel of Figure \ref{fig:cloud_sims} for different $N_{\text{Clouds}}$. The observed variable fractions are systematically lower when more clouds are simultaneously ingressing and egressing. The dashed, orange line shows that this is not simply due to the higher overall column densities, and therefore larger assumed measurement errors. The two fractions considering 9 clouds are in essentially perfect agreement, despite one being $\sim5$ times more absorbed than the other. Therefore these fractions are driven by a true difference in the variable behavior when more or less clouds are participating in the obscuring at any given time. 

This behavior is perhaps expected, as the higher $N_{\text{Clouds}}$ makes it more likely for the variability caused by an individual cloud to be opposed by variability in the other direction of another cloud. Large changes in $N_H$ would then require a significant fraction of the clouds to either line up or get out of the way simultaneously. The probability of this occurring decreases with the number of clouds. 

In the bottom left panel of Figure \ref{fig:cloud_sims}, the average values from Table \ref{tab:combos} \citep[][method]{NTA23} are shown as black points. We calculate the errors on these fractions as either the standard deviation between the model values or the binomial errors for each model added in quadrature. We show the larger of the two. The data for NGC\,1142 appear to prefer a moderate number of clouds ($\sim$9--12) participating in eclipses simultaneously, with $<6$ clouds apparently being ruled out. However, we stress the simplistic nature of this model and refrain from making strong conclusions based on it. Furthermore, we have not properly considered the errors on the simulated curves in the bottom left panel of Figure \ref{fig:cloud_sims}. Once errors are considered, it is unclear whether this method will remain powerful on a single source as suggested here. It may be that this measurement can only be used to obtain average properties on a population of sources. In this regard, a broader sample of heavily obscured AGN with multi-epoch NuSTAR and XMM-Newton observations would be necessary to test the validity of our assumptions. Presently, we only wish to convey the potential of using measurements such as these to constrain more realistic obscuration models.

\subsection{Qualitative Assessment of the Cloud Locations}

In addition to using the variable fraction to constrain the number of clouds, we attempt to constrain the location of the dominant obscuring material by considering the dependence of variability on the timescale between observations. The bottom right panel of Figure \ref{fig:cloud_sims} shows the difference in column density $\Delta N_H$ between pairs of observations at all timescales for three different cloud locations: the broad line region only ($10^{-3}\text{\,pc}\leq r\leq10^{-1}$\,pc), the torus only ($10^{-1}\text{\,pc}\leq r\leq10$\,pc), and both the broad line region and the torus ($10^{-3}\text{\,pc}\leq r\leq10$\,pc). The three lines in the figure indicate the 90$^{\rm th}$ percentile of the $\Delta N_H$ spread within each $\Delta t$ bin. The data were binned with 200 points per bin. Roughly, one would expect $\sim10$\,\% of the observed points to fall above these lines while the rest would be below. The spread of $\Delta N_H$ clearly increases with increasing timescale in all cases. However, it increases faster in the cases with clouds at smaller radii, because these clouds are assumed to be smaller, denser, and have higher orbital velocities.

We also plot the $\Delta N_H$ for all observation pairs of NGC\,1142 for the \texttt{UXCLUMPY} model. These were calculated from the $N_H$ values in Table \ref{tab:results_constant_gamma}. These data show the same trend as the simulated values. Only 2/36 points $(\sim 6\,\%)$ appear to be confidently above all three lines, which is to be expected since the lines represent the 90$^{\rm th}$ percentile. Unfortunately, there are not enough observations, and the uncertainties are too large to distinguish between the different scenarios. The largest predicted difference appears on timescales $\Delta t \sim 20-200$\,days, where there are only $\sim10$ pairs. The fact that none of the points have $\Delta N_H\gtrsim3\times10^{23}$\,cm$^{-2}$ could be interpreted as evidence against the BLR only scenario, however, with only 10 samples, the evidence is very weak and we make no claim from it. Therefore, we conclude that the data are consistent with all three simulated scenarios. Many more points, particularly with time separations $\Delta t \sim 20-200$\,days are needed to properly distinguish between the three scenarios of this model using this method. 

In any case, we emphasize that we do not aim to place physical constraints using this model. Rather, we intend to illustrate how these measurements can be used to constrain the location of the dominant obscuring material. In a future work, we will provide a more physically realistic model and a more thorough comparison to the available data. Additionally, we will summarize the results from all sources in our larger sample \citep[][]{pizzetti_multi-epoch_2022,NTA23,Pizzetti2025,NTA25,Sengupta25,Gianolli2026}.


\section{Summary} \label{sec:conclusions}

In this work, we have analyzed X-ray observations of NGC\,1142 from 9 different epochs spanning $\sim$2\,decades. We used three different physically-motivated torus models to describe the emission and constrain parameter values, the most important of which being the line-of-sight column density, $N_H$. All three models show variability in $N_H$, photon index, and luminosity on timescales $<$2\,years, and indicate a potential change from Compton-thin to Compton-thick and back again over the course of the 20 years of available data. Our main conclusions are the following:
\begin{enumerate}
	\item The $N_H$ is robustly constrained in all three models for most of the epochs. While the models occasionally have strong disagreements in photon index, these disagreements do not translate to large disagreements in $N_H$. More importantly, the variability in $N_H$ is even less affected since the measured trends are still in agreement, even with a slight systematic offset between the models.
	\item In other works by our group, the photon index was assumed to be constant across all epochs \citep[][]{marchesi_compton-thick_2022,pizzetti_multi-epoch_2022,NTA23,Pizzetti2025,NTA25,Sengupta25,Gianolli2026}. In this work, we followed the same procedure to allow our results to be easily compared to those other works. However, we also performed the analysis allowing photon index to vary. In doing so, we discovered that two of the observations prefer a much softer photon index relative to the other 7 observations. Also, during these observations, the luminosity was greatly increased following the softer-when-brighter behavior seen in other sources and AGN populations \citep[e.g.][]{Sobolewska2009,Caballero-Garcia2012,Soldi2014,Yang2015,Trakhtenbrot2017,Tortosa2024}. However, our main focus is on the $N_H$ variability. Importantly, despite the large photon index variability observed, none of our conclusions about $N_H$ variability change depending on whether photon index is tied or untied between epochs. This provides a strong confirmation to the $N_H$ variability determinations in previous works where photon index variability was not considered.
	\item The parameters describing the torus geometry are quite well constrained, compared to other sources in our sample. The \texttt{UXCLUMPY} results are in good agreement with the previous analysis on NGC\,1142 which used only 4 of the epochs available in this work \citep{Pizzetti2025}. To the extent that these models can be compared, \texttt{UXCLUMPY} and \texttt{XSKIRTor} are in good agreement with each other, showing similar inclination angles and $\sigma_{\text{Tor}}\sim C_f$ (although these should not necessarily be compared). The covering factor of \texttt{RXTorusD} is in agreement with \texttt{XSKIRTor} when the photon index is allowed to vary between epochs. The inclination angle of \texttt{RXTorusD}, however, appears to prefer a much more face-on inclination angle than the other two models. In any case, these values do not affect the $N_H$ determinations.
	\item Whether or not $N_H$ variability is detected in NGC\,1142 depends heavily on the number of observations considered. This dependence is similar to the dependence seen in populations of sources \citep[e.g.][]{NTA25,Cox26,Gianolli2026}. When only two epochs are considered, the probability of detecting variability is only $\sim$50\,\%. The probability increases to 100\,\% beyond 7 observations. 
	\item The spread in magnitude of the $N_H$ variability, $\Delta N_H$, increases as the timescale between epochs, $\Delta t$ increases. This too is a confirmation of trends seen in our previous works. That is, small changes in $N_H$ ($\Delta N_H\lesssim5\times10^{23}$\,cm$^{-2}$) are seen at all timescales, however, larger changes are only seen on timescales $\gtrsim$100\,days. This has previously been interpreted as evidence for the distant torus material contributing to the variability in $N_H$ along with BLR variability. However, the bottom panel of Figure \ref{fig:cloud_sims} shows that a BLR only scenario and a BLR+Torus scenario only result in slightly different expected observations. Therefore, given the amount of available data, we refrain from making the same claim here. Although, we do note that the data are very slightly more consistent with the BLR+Torus scenario. 
	\item A simple model to simulate `obscuration curves' was developed. These simulated curves were subjected to the same analysis as the NGC\,1142 data and the results were compared. We find that in order to explain the observed variability in NGC\,1142, many clouds ($\gtrsim6$) are likely required to be simultaneously ingressing and egressing the line of sight at all times. We show how the probability of observing variability in a source as a function of the number of observations can be used to constrain this property. Furthermore, the spread of $\Delta N_H$ as a function of $\Delta t$ can also be used to constrain the location of material primarily responsible for the observed $N_H$ variability. Unfortunately, the NGC\,1142 data are not constrained well enough to make a firm conclusion on this. 
\end{enumerate}

In a future work, we will refine the simple model introduced here and present the results of the full sample ($\gtrsim$40\,sources). We will also explore better ways to compare the simulated and observed variability patterns and use these comparisons to better constrain properties of the X-ray obscuring material in AGN.

\section{Acknowledgments}

I.C. acknowledges support under grant AR4-25009X and 80NSSC24K1850. V.E.G. acknowledges funding under NASA contract 80NSSC24K1403. We have made extensive use of the \texttt{NumPy} \citep{numpy} and \texttt{Pandas} \citep{pandas} packages in the \texttt{Python3} \citep{python} programming language.

I.C. would like it noted that the use of generative A`I' was deliberately avoided in this work. Of course, it is likely that some sources studied during the process of creating this work contained material generated by an LLM. However, I.C. is unaware of any specific instances of consuming non-human generated text or `ideas' and regrets that this is the state of the internet and scientific literature.

\bibliography{references}{}
\bibliographystyle{aasjournal}

\end{document}